\documentclass[aps,prl,preprint,groupedaddress]{revtex4}
\usepackage{mathrsfs,txfonts,graphicx}
\catcode`@=11
\@ifundefined{pdfoutput}{\usepackage[dvips,bookmarks]{hyperref}}{\usepackage{hyperref}}
\catcode`@=12
\hypersetup{colorlinks,bookmarksopen,bookmarksnumbered,citecolor=blue,pdfstartview=FitH}
\def\hhref#1{\href{http://arxiv.org/abs/hep-th/#1}{hep-th/#1}}
\def\phref#1{\href{http://arxiv.org/abs/hep-ph/#1}{hep-ph/#1}}
\def\mhref#1{\href{mailto:#1}{#1}}

\begin{document}
\preprint{YITP-SB-06-29}
\title{Worldline Green Functions for Arbitrary Feynman Diagrams}
\author{Peng Dai\footnotemark \footnotetext{Electronic address: \mhref{pdai@grad.physics.sunysb.edu}}
and Warren Siegel\footnotemark \footnotetext{Electronic address:
\mhref{siegel@insti.physics.sunysb.edu}}}
 \affiliation{C. N. Yang Institute for Theoretical Physics, Stony Brook University, Stony Brook, NY 11790-3840}
\date{August 9, 2006; revised: February 1, 2007}

\begin{abstract}
We propose a general method to obtain the scalar worldline Green
function on an arbitrary 1D topological space, with which the first-quantized method of evaluating 1-loop Feynman diagrams can be
generalized to calculate arbitrary ones. The electric analog of
the worldline Green function problem is found and a compact
expression for the worldline Green function is given, which has
similar structure to the 2D bosonic Green function of the closed
bosonic string.
\end{abstract}
\maketitle

\section{I. Introduction}

The first-quantized method for the calculation of particle
scattering amplitudes was suggested and used by Feynman
\cite{Feynman:1950ir}, Nambu \cite{Nambu:1950rs} and Schwinger
\cite{Schwinger:1951nm} in as early as the 1950's. However, it was not
developed much until string theory was invented and used
first-quantization as a key method to calculate the string
scattering amplitude \cite{Susskind:1970qz}. Then, how to apply this
method back to particles was studied in detail. From a certain limit
of string theory, Bern and Kosower obtained a set of first-quantized
rules for calculating the scattering amplitudes in Yang-Mills theory
at the 1-loop level \cite{Bern:1991aq}. This new method was shown to be
equivalent to the ordinary second-quantized method and much more
efficient when calculating 1-loop gluon-gluon scattering
\cite{Bern:1991an}. Later Strassler gave an alternative derivation
of the same rules directly from the first-quantized formalism of the
field theory \cite{Strassler:1992zr}.

The generalization to the rules for scalar theory at multi-loop
level was then studied by Schmidt and Schubert \cite{Schmidt:1994zj}
and later by Roland and Sato \cite{Roland:1996im}. Green functions
on multi-loop vacuum diagrams were obtained by considering these
diagrams as one-loop diagrams with insertions of propagators. The
Green function on any vacuum diagram containing a ``Hamiltonian
circuit" could be found by this method.

A natural hope of further generalization is to find the worldline
Green function on an arbitrary one-dimensional topology, without the
limitation that this topology must be one-loop with insertions. In
this paper, we give such a general method to obtain the Green
function for scalar field theory at arbitrary multi-loop level. We
show that the electric circuit can be an analog in solving this
problem.

On the other hand, Mathews \cite{Mathews:1959} and Bjorken
\cite{Bjorken:1979dk}, from the second-quantized method (usual field
theory), gave a method with the electric circuit analogy to evaluate
Feynman diagrams at arbitrary loop level. Their result of the Feynman
parameter integral representation of scattering amplitudes was, in
principle, the most general result, but because of the limitation of
second-quantization, diagrams with the same topology except for
different number or placement of the external lines were treated
separately, and the analogy between the kinetic quantities on the
Feynman diagram and the electric quantities on the circuit was not
completely clear.

Fairlie and Nielsen generalized Mathews and Bjorken's circuit analog
method to strings, and obtained the Veneziano amplitude and 1-loop
string amplitude \cite{Fairlie:1970tc}. Although they didn't use the
term ``Green function", they in fact obtained the expression of the
Green function on the disk and annulus worldsheets of the bosonic string
with the help of the 2D electric circuit analogy.

These attempts indicate that the problem of solving for the Green function
on a certain topological space and the problem of solving a circuit
may be related. Indeed, we show that there is an exact analogy
between the two kinds of problems in the 1D case.

In this paper, we first give an introduction to the first-quantized
formalism in Section II. In Section III, we find
that the
differential equation the Green function should satisfy can be solved by an analogous method for solving the electric
circuit. We give a complete analogy among the problems of finding
the Green function, the static electric field and the electric circuit. In
section IV, we derive a general method to solve the electric circuit
and give a compact expression of the Green function,
    \[
        \tilde{G}(\tau,\tau')=-\frac{1}{2} s + \frac{1}{2} {\mathbf{v}}^{\mbox{T}}\Omega ^{-1}{\mathbf{v}}
    \]
where the scalar $s$, vector $\mathbf{v}$, and matrix $\Omega$ are
quantities depending only on the topological and geometrical
properties of the 1D space and the position of the external sources
$\tau$ and $\tau'$. This expression is similar to that of the
bosonic string \cite{D'Hoker:1988ta}. In Section V, a calculation of
the vacuum bubble amplitude is given to complete the discussion. The
method is summarized in section VI. Examples of Green functions on
topologies at the tree, 1-loop and 2-loop levels are given in
Section VII. Section VIII contains the conclusions.

\section{II. First-quantization}

The first-quantized path integral on the ``worldline" of any given
topology (graph) for a scalar particle interacting with a background potential
can be written as
   \begin{equation}\label{eq1}
      \mathscr {A}\left(\mbox{M}\right) = \int
      \frac{\mathscr{D}e\mathscr{D}X}{V_{\mbox{rep}}} \exp \left\{ -
      \int_{\mbox{M}} d \tau \left[ \frac{1}{2} \left( e^{-1} \dot{X}^\mu
      \dot{X}_\mu + em^2 \right) + e V(X) \right] \right\}
   \end{equation}
where $\mbox{M}$ means the particular topology of the worldline
being considered and the potential $V(X)$ is specialized to a
background consisting of a set of plane waves for our purpose to
calculate the scattering amplitude
    \[
       V \left( X \right)=\sum\limits_{i=1}^N {e^{ik_i \cdot X\left(
       {\tau _i } \right)}}
    \]
$V_{\mbox{rep}}$ in (\ref{eq1}) is the volume of the
reparametrization group. The reparametrization symmetry of the
worldline must be fixed to avoid overcounting.

The path integral (\ref{eq1}) gives the sum of all graphs of a given
topology with an arbitrary number of external lines. This can be seen
by expanding the potential exponential. If we consider only graphs
with a certain number of external lines, the amplitude can be written
as
    \begin{equation}\label{eq2}
        \mathscr {A}\left(\mbox{M},N\right) =\int {\frac{\mathscr{D}e\mathscr{D}X}
        {V_{\mbox{rep}}}\exp \left[{-\frac{1}{2}\int_{\mbox{M}} {d\tau \left(
        {e^{-1}\dot {X}^\mu \dot {X}_\mu +em^2} \right)} } \right]\cdot
        \prod\limits_{i=1}^N {\int_{\mbox{M}} { d\tau_i e \left[ e^{ik_i \cdot
        X\left( {\tau _i } \right)} \right] } } }
    \end{equation}
where $N$ is the number of external lines.

To fix the reparametrization symmetry, we can simply set the
vielbein $e$ to $1$. However, by doing this, we have left some part
of the symmetry unfixed (the ``Killing group", like the conformal
Killing group in string theory), as well as over-fixed some
non-symmetric transformation (the ``moduli space", also like in
string theory). To repair these mismatches by hand, we add integrals
over the moduli space and take off some of the integrals over the
topology. The general form of the amplitude (\ref{eq2}) after fixing
the reparametrization symmetry is (up to a constant factor from
possible fixing of the discrete symmetry which arises due to the
indistinguishable internal lines)
    \begin{equation}\label{eq3}
        \mathscr {A}( \mbox{M},N)=\prod\limits_{a=1}^\mu
        \int_{\mbox{F}_a } dT_a \int \mathscr{D}{\rm X}\exp \left[
        -\frac12\int_{\mbox{M}} d\tau \left( \dot X^\mu \dot X_\mu
        +m^2 \right) \right]
        \int_{\mbox{M}} \left( \prod\limits_{c\notin \mbox{C}}d\tau _c \right)
        \prod_i e^{ik_i \cdot X_i( \tau _i)}
    \end{equation}
where $C$ denotes the external lines whose position has to be fixed
to fix the residual symmetry, and in the particle case the moduli
$T_a$ are just the ``proper times" represented by the lengths of the
edges in the graph of topology $\mbox{M}$. Further evaluation by the
usual method of 1D field theory gives the following expression (with
the delta function produced by the zero mode integral suppressed
since it trivially enforces the total momentum conservation in the
calculation of the scattering amplitude),
    \begin{equation}\label{eq4}
        \mathscr {A}( \mbox{M},N)=\prod\limits_{a=1}^\mu
        \int_{\mbox{F}_a } dT_a\, \mathscr{V}_{\mbox{M}}( T_a)
        \int_{\mbox{M}} \left( \prod\limits_{c\notin \mbox{C}}d\tau _c \right)
        \exp \left[ -\frac12\sum\limits_{i,j} k_i \cdot k_j G_{\mbox{M}}(
        \tau _i ,\tau _j ) \right]
    \end{equation}
where $\mathscr{V}_{\mbox{M}}\left( T_a \right)$ is the amplitude of
the vacuum bubble diagram,
    \begin{equation}\label{vacuum bubble}
        \mathscr{V}_{\mbox{M}}\left( T_a \right) = \int {\mathscr{D}{\rm X}\exp \left[
        {-\frac{1}{2}\int_{\mbox{M}} {d\tau \left( {\dot {X}^\mu \dot {X}_\mu
        +m^2} \right)} } \right] }
    \end{equation}
and $G_{\mbox{M}}$ is the Green function which satisfies the
following differential equation on the worldline of topology M:
    \begin{equation}\label{eq5}
        \ddot {G}_{\mbox{M}} \left( {\tau ,{\tau }'} \right)=-\delta \left(
        {\tau -{\tau }'} \right)+\rho
    \end{equation}
where $\rho$ is a constant, of which the integral over the whole 1D
space gives $1$, i.e., the inverse of the total volume (length) of
the 1D space
    \[
        \rho = \frac {1} {\int_{\mbox{M}} {d\tau }}
    \]
This constant is required by the compactness of the
worldline space.

\begin{figure}
    \includegraphics{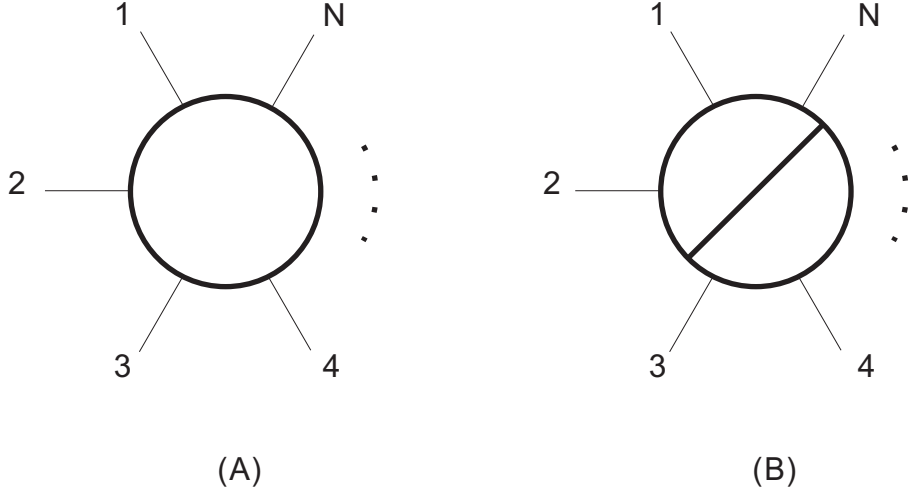}
    \caption{\label{fig1}
        (A) One-loop diagram with $N$ external lines.
        The topology of a circle has one modulus,
        the circumference $T$ of the loop.
        Also there is one residual symmetry, which has
        to be fixed by taking off one of the proper-time integrals.
        (B) Two-loop diagrams with $N$ external lines.
        The topology of this kind has three moduli,
        the lengths $T_1$, $T_2$ and $T_3$ of the three edges.
        And there is no unfixed residual symmetry.}
\end{figure}

The scattering amplitude (\ref{eq4}) is a general expression for a
worldline of any topology with an arbitrary number of external lines.
The form of the vacuum bubble amplitude and Green function depends
on the topology. For example, the amplitude of the
one-loop 1PI graph with $N$ external lines (Fig.\ \ref{fig1}(A)) is
    \[
        \mathscr {A}\left( {\bigcirc , N} \right)=\int_0^\infty dT\,
        \mathscr{V}_{\bigcirc}
        {\int_\bigcirc {\left( \prod\limits_{c=1}^{N-1}{d\tau_c} \right)
        \exp\left[ {-\frac{1}{2}\sum\limits_{i,j} {k_i \cdot k_j
        G_{\bigcirc}
        \left( {\tau _i ,\tau _j } \right)} } \right]} }
    \]
where
    \[
        \mathscr{V}_{\bigcirc}\left( T \right)
        =  \exp \left( -\frac{1}{2} T m^2 \right) T ^ {-D/2}
    \]
and
    \[
        G_\bigcirc \left( {\tau ,{\tau }'} \right)=-\frac{1}{2}\left[ {\left| {\tau
        -{\tau }'} \right|-\frac{\left( {\tau -{\tau }'} \right)^2}{T}}\right]
    \]
Another example is the amplitude of the two-loop 1PI graph with $N$
external lines (Fig.\ \ref{fig1}(B)),
    \[
        \mathscr {A}\left( {\ominus , N} \right)=\prod\limits_{a=1}^3
        {\int_0^\infty {dT_a } } \,
        \mathscr{V}_{\ominus}\left(T_1,T_2,T_3\right)
        {\int_{\ominus} {\left( \prod\limits_{c=1}^N{d\tau_c} \right)
        \exp \left[ {-\frac{1}{2}\sum\limits_{i,j}
        {k_i \cdot k_j G_{\ominus} \left( {\tau _i ,\tau _j } \right)} } \right]} }
    \]
where $\mathscr{V}_{\ominus}$ and $G_{\ominus}$ will be determined
in the following sections.

\section{III. Green function}

We note that the differential equation (\ref{eq5}) is just the
Poisson equation the electric potential should satisfy when there is
a unit positive charge at ${\tau}'$ and a constant negative charge
density of magnitude $\rho$ over the whole space. This suggests to
us to consider the corresponding static electric problem where the
Green function is just the electric potential at $\tau$ of the above
setup.

\begin{figure}
    \includegraphics{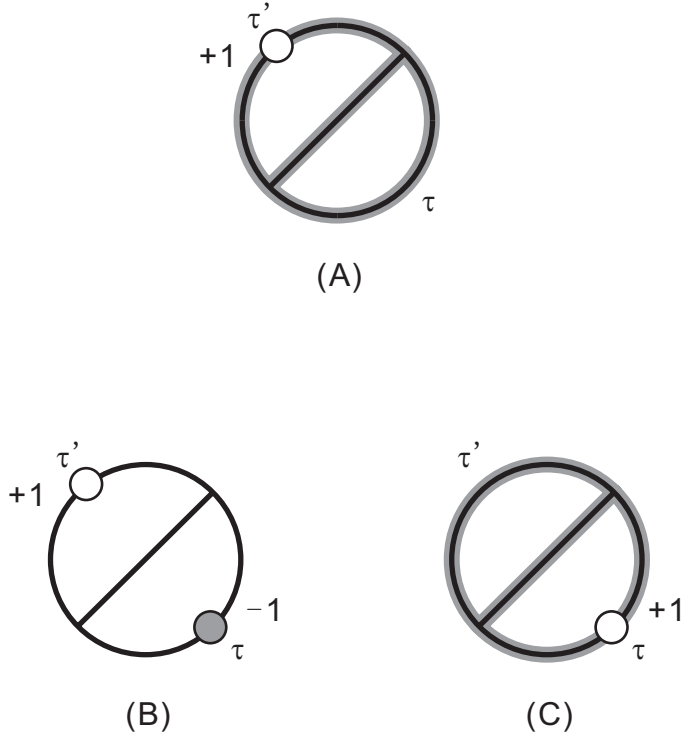}
    \caption{\label{fig_sum}
        (A) Two-loop topological space with a unit positive charge at $\tau'$
        and a unit negative charge uniformly distributed over the whole
        space. (B) Two-loop topological space with a unit positive
        charge at $\tau'$ and a unit negative charge at
        $\tau$. (C) Two-loop topological space with a unit positive
        charge at $\tau$ and a unit negative charge uniformly distributed
        over the whole space.}
\end{figure}

To demonstrate the general solution of the Poisson equation, we
solve for the Green function on the two-loop worldline as an
example. Consider the 1D topological space as shown in Fig.\
\ref{fig_sum}(A) with a unit positive charge at $\tau'$ and a unit
negative charge uniformly distributed over the whole space.
According to the above argument, the Green function $G \left( \tau ,
\tau' \right)$ is the electric potential at $\tau$. Now we can use
Gauss' law and single-valuedness of the potential to write down
equations and solve for the expression of the potential. The
potential indeed gives a right form for the Green function, but it
contains many terms that will be canceled out when calculating the
scattering amplitude using equation (\ref{eq4}), and hence can be
further simplified.

Note that the setup of the static electric problem in Fig.\
\ref{fig_sum}(A) can be regarded as the superposition of the 2 setups in
Fig.\ \ref{fig_sum}(B)(C).
Let $G\left( \tau , \tau' \right)$ denote the potential at $\tau$ of
the setup shown in \ref{fig_sum}(A), and $\bar {G} \left( \tau ,
\tau' \right)$ denote the potential at $\tau$ in Fig.\
\ref{fig_sum}(B). The potential at $\tau$ in Fig.\ \ref{fig_sum}(C) is
then $G\left( \tau , \tau \right)$. Thus
    \[
        G\left( \tau , \tau' \right) = \bar {G} \left( \tau ,
        \tau' \right) + G\left( \tau , \tau \right)
    \]
We further define $\tilde {G} \left( \tau , \tau' \right)$ as the
symmetric part of $\bar {G} \left( \tau , \tau' \right)$, i.e.,
    \begin{equation} \label{G_tilde}
        \tilde {G} \left( \tau , \tau' \right) \equiv
        \frac{1}{2} \left[ \bar {G} \left( \tau , \tau' \right)
        + \bar {G} \left( \tau , \tau' \right) \right]
        =\left[ \frac{1}{2} G \left( \tau , \tau' \right) + \frac{1}{2} G \left( \tau' , \tau
        \right) \right]
        -\frac{1}{2} G \left( \tau , \tau \right) -\frac{1}{2} G \left( \tau' , \tau' \right)
    \end{equation}
If we use $\tilde {G} \left( \tau , \tau' \right)$ instead of $G
\left( \tau , \tau' \right)$ in equation (\ref{eq4}), the sum can be
rewritten as
\begin{eqnarray*}
            -\frac{1}{2}\sum\limits_{i,j} {k_i \cdot k_j \tilde{G}\left( {\tau _i ,\tau _j }
            \right)}
            & = & -\frac{1}{2}\sum\limits_{i,j} {k_i \cdot k_j \left\{
            \frac{1}{2} \left[
            G \left( {\tau _i ,\tau _j} \right)+ G \left( {\tau _j ,\tau _i}
            \right) - G \left( {\tau _i ,\tau _i} \right) -G \left( {\tau _j ,\tau _j} \right)
            \right]\right\}}\\
            & = & -\frac{1}{2}\sum\limits_{i,j} {k_i \cdot k_j \left\{
            \frac{1}{2} \left[
            G \left( {\tau _i ,\tau _j} \right)+ G \left( {\tau _j ,\tau _i}
            \right) \right]\right\}}\\
            & = & -\frac{1}{2}\sum\limits_{i,j} {k_i \cdot k_j G\left( {\tau _i ,\tau _j }
            \right)}
\end{eqnarray*}
We have used conservation of momentum $\sum{k_i}=0$ in the second
step and rearranged the summands in the third step. This shows that
using $\tilde {G} \left( \tau , \tau' \right)$ in equation
(\ref{eq4}) is equivalent to using $G \left( \tau , \tau' \right)$.

\begin{figure}
\includegraphics{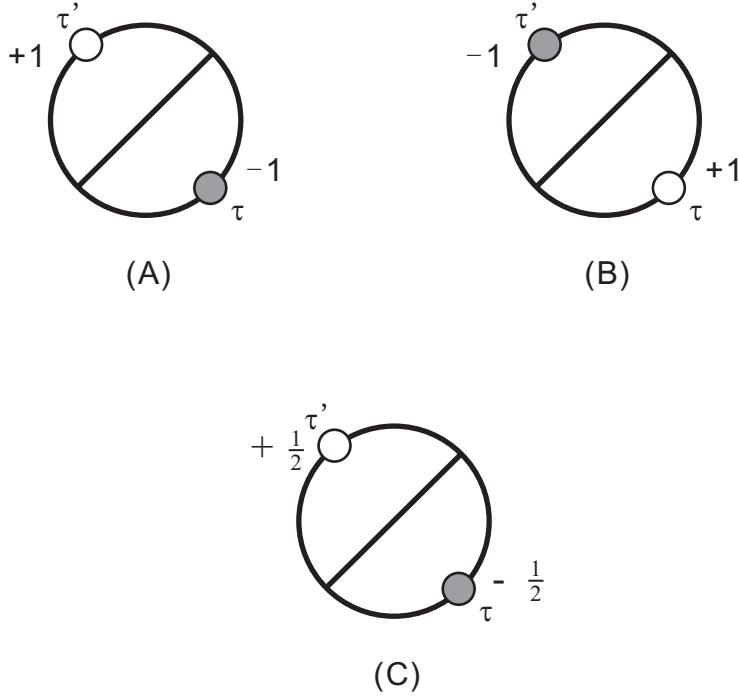}
\caption{\label{fig_symmetry}
    (A) Two-loop topological space with a unit positive
    charge at $\tau'$ and a unit negative charge at
    $\tau$. (B) Two-loop topological space with a unit negative
    charge at $\tau'$ and a unit positive charge at
    $\tau$. (C) Two-loop topological space with a half unit positive
    charge at $\tau'$ and a half unit negative charge at
    $\tau$.}
\end{figure}

Note that same procedure is usually applied to construct the Green
function for bosonic strings \cite{D'Hoker:1988ta},
    \[
        \tilde {G} \left( z , w \right)= G \left( z , w \right) - \frac{1}{2} G \left( z , z \right) -\frac{1}{2} G \left( w, w \right)
    \]
which is similar to equation (\ref{G_tilde}).

Now we have to look for an electric field analog of $\tilde {G}
\left( \tau , \tau' \right)$. Since $\bar {G} \left( \tau , \tau'
\right)$ is the electric potential at $\tau$ of the setup shown in
Fig.\ \ref{fig_symmetry}(A), it can be written as the potential at
$\tau'$ plus the potential difference from $\tau$ to $\tau'$. And
$\bar {G} \left( \tau' , \tau \right)$ is the potential at $\tau'$
of the setup shown in Fig.\ \ref{fig_symmetry}(B). The sum of $\bar
{G} \left( \tau , \tau' \right)$ and $\bar {G} \left( \tau' , \tau
\right)$ just gives the potential difference from $\tau$ to $\tau'$
of Fig.\ \ref{fig_symmetry}(A) because the potential at $\tau'$ of Fig.\
\ref{fig_symmetry}(A) cancels the potential at $\tau'$ of Fig.\nobreak\ \ref{fig_symmetry}(B). $\tilde {G} \left( \tau , \tau' \right)$ is
half that potential difference, and therefore is just the potential
difference from $\tau$ to $\tau'$ of Fig.\ \ref{fig_symmetry}(C).

It is now clear that, to write down the expression of the scattering
amplitude, we only have to know the symmetric Green function $\tilde
{G}$, and use the following formula:
\begin{equation}\label{eq10}
        \mathscr {A}( \mbox{M},N)=\prod_{a=1}^\mu
        \int_{\mbox{F}_a}  dT_a\, \mathscr{V}_{\mbox{M}} \prod_{c\notin \mbox{C}}
        \int_{\mbox{M}} d\tau _c \exp \left[
        -\sum_{i<j} k_i \cdot k_j \tilde G( \tau _i ,\tau _j) \right]
    \end{equation}
This simplifies the expression of the Green function.

\begin{figure}
\includegraphics{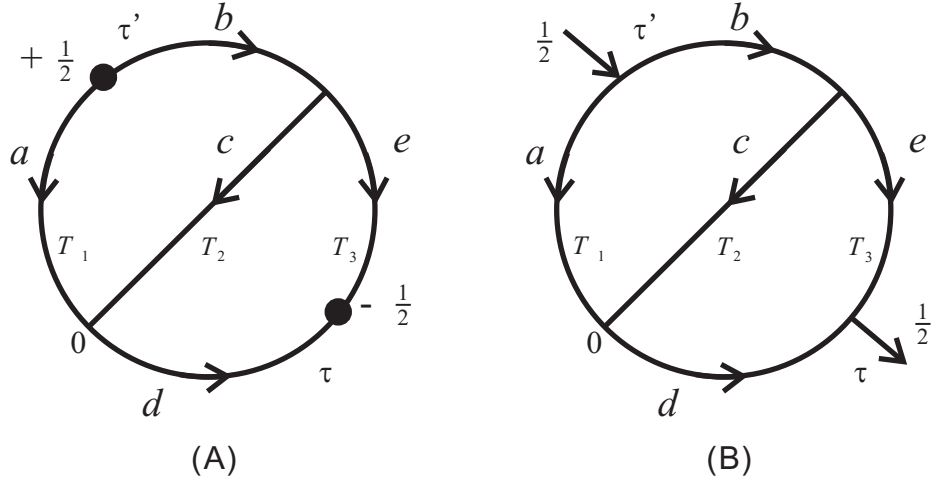}
\caption{\label{fig3}
    (A) Two-loop topological space with a half unit positive
    charge at $\tau'$ and a half unit negative charge at $\tau$.
    The lengths of the three arcs are $T_1$, $T_2$,
    $T_3$. $\tau'$ and $\tau$ are respectively on $T_1$ and $T_3$
    and denote the lengths from the origin.
    The magnitudes of the electric field on each part of the space
    are denoted by $a$ - $e$ and the directions are
    chosen arbitrarily. (B) Two-loop circuit with a half unit
    current input at $\tau'$ and withdrawn at $\tau$.
    The resistances of the three arcs are $T_1$, $T_2$,
    $T_3$. $\tau'$ and $\tau$ are respectively on $T_1$ and $T_3$
    and denote the resistance from the origin. The currents on the parts of the circuit are denoted by $a$ - $e$
    and the directions are chosen arbitrarily.}
\end{figure}

Now that $\tilde {G} ( \tau , \tau')$ is the potential difference
from $\tau$ to $\tau'$ in Fig.\ \ref{fig_symmetry}(C), we can apply
Gauss' law (of 1D space) and single-valuedness of the potential to
write down the equations the Green function (electric potential
$\Phi$) and its first derivative (electric field $E$) should
satisfy. We assume the direction and value of $E$ as shown in Fig.\
\ref{fig3}(A) and $\tau'$ and $\tau$ are respectively on $T_1$ and
$T_3$. According to Gauss' law, we have the following equations:
\begin{eqnarray*}
        a+b&=&+\frac{1}{2}\\
        -a-c+d&=&0\\
        -b+c+e&=&0\\
        -d-e&=&-\frac{1}{2}
\end{eqnarray*}
The single-valuedness of the potential requires
\begin{eqnarray*}
        cT_2+b(T_1-\tau')&=&a\tau'\\
        cT_2+d\tau&=&e(T_3-\tau)
\end{eqnarray*}
If we regard the electric field $E$ as the current $I$, the electric
potential $\Phi$ as the voltage $V$, and the length on the 1D space
$l$ as the resistance $r$, these equations are just the Kirchhoff
equations of a circuit of the same shape as the worldline and with
half unit current going into the circuit at $\tau'$ and half unit
current coming out at $\tau$, as shown in Fig.\ \ref{fig3}(B). It is
easy to see that this equivalence between 1D static electric field
and circuit is valid for an arbitrary 1D topological space.  We have
the following relations for 1D (note that there is no
cross-sectional area in the 1D case):
\begin{eqnarray*}
        \sigma E &=& I\\
        \rho l &=& r\\
        \Phi &=& V
\end{eqnarray*}
where $\sigma$ and $\rho$ ($=1/\sigma$) are respectively the 1D
conductivity and resistivity, and the relation $El=\Phi$ is
equivalent to Ohm's law $Ir=V$. In addition, Gauss' law is
equivalent to Kirchhoff's current law, and the single-valuedness of
the potential is equivalent to Kirchhoff's voltage law. Thus the two
problems are indeed equivalent. The Green function $\tilde{G} (\tau
, \tau')$ on a particular 1D topological space $\mbox{M}$ can be
understood as the voltage difference from $\tau$ to $\tau'$ when a
half unit current is input into the ``circuit" (1D space) at $\tau'$
and withdrawn at $\tau$.

Above we have shown that in the 1D case the problem of solving for the
first-quantized particle Green function, the potential difference in a
static electric field and the voltage difference in a circuit are
analogous. Before we solve the circuit problem for the Green
function, it is interesting to give a complete analogy among the
quantities in these three kinds of problems (Table \ref{tab1}).

\begin{table}[!h]
\tabcolsep 0pt \caption{\label{tab1}Analogy among quantities in
three kinds of problems} \vspace*{-12pt}
\begin{center}
\def\temptablewidth{0.8\textwidth}
{\rule{\temptablewidth}{1pt}}
\begin{tabular*}{\temptablewidth}{@{\extracolsep{\fill}}cccccc}
       Particle & Static electric field & Electric circuit \\
       \hline
       Green function $\tilde {G}$ & potential difference $\Delta \Phi$
       & voltage difference $\Delta V$ \\
       position $x$ & electric potential $\Phi$ & voltage $V$ \\
       momentum $p$ & electric field $E$ & current $I$ \\
       proper time $\tau$ & length $l$ & resistance $r$ \\
       action $S$ & energy $U$ & power $P$ \\
       external force $F$ & electric charge $Q$ & emf $\mathscr{E}$
       \end{tabular*}
       {\rule{\temptablewidth}{1pt}}
\end{center}
\end{table}

\section{IV. Solving the circuit}

The next thing we have to do is to find a general expression for the
Green function, i.e., a method to obtain the voltage difference
described above. A formula has already been given in graph theory
(see, e.g., \cite{Bollobas:1979bq}). Here we summarize this result and develop it
into a formula which is similar to the known form of the Green
function on the 2D worldsheet of bosonic string theory
\cite{D'Hoker:1988ta}.

The voltage difference from $\tau$ to $\tau'$ can be obtained by the
following steps:

(1) Connect $\tau$ and $\tau'$ by a wire with zero resistance.

(2) The resulting graph has vertices $V={v_1,...,v_n}$ and edges
$E={e_1,...e_{m-1},e_m}$. Denote by $e_m$ the edge (wire) we have
just added in. Set the directions of all the edges arbitrarily.

(3) Assign voltage, current and resistance on each edge; they can be
written in the form of vectors:
\begin{eqnarray*}
        \mathbf{U} & = &(U_1,U_2,...,U_m)^{\mbox{T}}\\
        \mathbf{I} & = &(I_1,I_2,...,I_m)^{\mbox{T}}\\
        \mathbf{r} & = & (r_1,r_2,...,r_m)^{\mbox{T}}
\end{eqnarray*}
Also define the external-force-driven voltage $\mathbf{u}$, which
has, in our case, only one non-zero component (the last one). Assume
it has magnitude $1$:
    \begin{equation} \label{eq6}
        \mathbf{u}=(u_1,u_2,...,u_m)^{\mbox{T}}=( 0,0,...,1 )^{\mbox{T}}
    \end{equation}

(4) Find all the ``independent" loops in the graph:  There should be
$(m-n+1)$. These loops can be identified by the following method:
Choose an arbitrary spanning tree (a connected subgraph that
contains all the vertices and is a tree) of the graph. There are
always $(m-n+1)$ chords (the edges not belonging to the spanning
tree). Adding a chord to the spanning tree will generate a one-loop
graph. Thus each chord gives a loop in the graph, and all the loops
obtained by this way are independent of each other. Therefore there
are $(m-n+1)$ loops. Assign an arbitrary direction to each loop.

(5) Define matrices $\mathcal{B}$, $\mathcal{C}$ and $\mathcal{R}$ as
follows:
{\jot=12pt
\begin{eqnarray*}
       {\mathcal{B}_{ij} } & = &
       \cases{
            1 & if $v_i$ is the initial vertex of $e_j$ \cr
            -1 & if $v_i$ is the terminal vertex of $e_j$ \cr
            0 & otherwise \cr} \\
       {\mathcal{C}_{ij} } & = &
       \cases{
            1 & if $e_i$ is in the same direction of the loop $l_j$ \cr
            -1 & if $e_i$ is in the opposite direction of the loop $l_j$ \cr
            0 & otherwise \cr} \\
        {\mathcal{R}_{ij} } & = &
       \cases{
            r_i & $i=j$ \cr
            0 & $i \neq j$ \cr}
\end{eqnarray*}
}

(6) With the above definitions, we can write down Kirchhoff's
current law, Kirchhoff's voltage law and Ohm's law in compact forms
as follows:
\begin{eqnarray*}
        \mathcal{B}\mathbf{I}&=&\mathbf{0}\\
        \mathcal{C}^{\mbox{T}}\mathbf{U}&=&\mathbf{0}\\
        \mathbf{U}&=&\mathcal{R}\mathbf{I}+\mathbf{u}
\end{eqnarray*}
The solution to the current on each edge is
    \[
        \mathbf{I} = -\mathcal{C}\left( {\mathcal{C}^{\mbox{T}}\mathcal{RC}}
        \right)^{-1}\mathcal{C}^{\mbox{T}}\mathbf{u}
    \]

(7) The total resistance between $\tau$ and $\tau'$ (excluding the
added [last] edge) is then minus the voltage on the last edge
divided by the current on that edge, i.e.,
    \[
        R(\tau,\tau') = - \frac{U_m}{I_m} = - \frac{u_m}{I_m} = - \frac{1}{I_m}
    \]
The Green function is then minus this resistance times the current,
one half, since it is the voltage difference from $\tau$ to $\tau'$,
    \begin{equation} \label{eq7}
        \tilde G(\tau,\tau') = - \frac{1}{2} R(\tau,\tau') = \frac{1}{2I_m}
        = - \frac{1}{2\mathbf{u}^{\mbox{T}} \mathcal{C}
        \left( {\mathcal{C}^{\mbox{T}}\mathcal{RC}} \right)^{-1}
        \mathcal{C}^{\mbox{T}} \mathbf{u}}
    \end{equation}
where the last step comes from extracting the last component of
$\mathbf{I}$ by using the vector $\mathbf{u}$ defined in equation
(\ref{eq6}).

We can further simplify equation (\ref{eq7}), by considering the
physical meaning of each part of the denominator:

(1) $\mathcal{C}^{\mbox{T}}\mathbf{u}$ and
${\mathbf{u}}^{\mbox{T}}\mathcal{C}$: Since $\mathbf{u}=\left(
{0,...,0,1} \right)^{\mbox{T}}$, $\mathcal{C}^{\mbox{T}}\mathbf{u}$
is an $\left( {m-n+1} \right)$-component column vector whose $i$th
component is the direction of the last ($m$th) edge on the $i$th
loop ($1$ for ``same", $-1$ for ``opposite", and $0$ for ``not on
the loop"). If we appropriately choose our independent loops, we can
achieve that the $m$th edge is only on the $\left( {m-n+1}
\right)$th (last) loop and has the same direction as this loop. This
is always achievable in steps: (a) Choose the spanning tree in such
a way that the $m$th edge doesn't belong to the spanning tree, i.e.,
is a chord. (b) Define the loop generated by adding the $m$th edge
to the spanning tree to be the $\left( {m-n+1} \right)$th loop. (c)
Define the direction of the $\left( {m-n+1} \right)$th loop to be
the same as the $m$th edge. By doing so, we find that
$\mathcal{C}^{\mbox{T}}{\rm {\bf u}}$ is just a $\left( {m-n+1}
\right)$ component column vector with the last component
non-vanishing and of value $1$.
    \[
        \mathcal{C}^{\mbox{T}}\mathbf{u}=\left( {0,...,0,1} \right)^{\mbox{T}}
    \]
And $\mathbf{u}^{\mbox{T}}\mathcal{C}$ is the transpose of
$\mathcal{C}^{\mbox{T}}\mathbf{u}$. Define for convenience
    \[
        \mathbf{P} \equiv \mathcal{C}^{\mbox{T}}\mathbf{u}
    \]
Note that $\mathbf{P}^{\mbox{T}}\mathcal{M}{\mathbf{P}}$ gives the
$\left[ {\left( {m-n+1} \right),\left( {m-n+1} \right)} \right]$
component of any matrix $\mathcal{M}$ with dimension $\left( {m-n+1}
\right)\times \left( {m-n+1} \right)$.

(2) $\mathcal{C}^{\mbox{T}}\mathcal{RC}$:
$\mathcal{C}^{\mbox{T}}\mathcal{RC}$ is an $\left( {m-n+1}
\right)\times \left( {m-n+1} \right)$ matrix. The components can be
interpreted as the sum of the ``signed" resistances,
    \[
        \left( {\mathcal{C}^{\mbox{T}}\mathcal{RC}} \right)_{ij}
        =\sum\limits_{k\,\in\,\rm{all\,edges}} {f\left( {k,i,j} \right)r_k }
    \]
where $r_k$ is the resistance on $k$th edge and $f$ is the ``sign":
    \begin{equation}\label{eq9}
        f(n,i,j) = \cases{
        1 & $n$th edge is on both loops $i$ and $j$ with
        the same orientation \cr
        -1 & $n$th edge is on both loops $i$ and $j$ with
        opposite orientations \cr
        0 & $n$th edge is not on both loops $i$ and $j$ \cr}
    \end{equation}
We define for convenience
    \begin{equation}\label{convenience}
        \mathcal{M} \equiv \pmatrix{
        \Omega & {\mathbf{v}} \cr
        {\mathbf{v}^{\mbox{T}}} & s \cr}
        \equiv \mathcal{C}^{\mbox{T}}\mathcal{RC}
    \end{equation}
where $\Omega$ is an $\left( {m-n} \right)\times \left( {m-n}
\right)$ matrix, $\mathbf{v}$ is an $\left( {m-n} \right)$-component
vector, and $s$ is a scalar.

Now the formula for the Green function (\ref{eq7}) becomes
    \[
        \tilde G = -\frac {1} { 2 \mathbf{P}^{\mbox{T}} \mathcal{M}^{-1} \mathbf{P} }
    \]
Since ${\mathbf{P}}^{\mbox{T}}\mathcal{M}^{-1}{\mathbf{P}}$ is just
the $\left[ {\left( {m-n+1} \right),\left( {m-n+1} \right)} \right]$
component of $\mathcal{M}^{-1}$, we have
    \[
        {\mathbf{P}}^{\mbox{T}}\mathcal{M}^{-1}{\mathbf{P}}=\frac{\det \Omega}{\det \mathcal{M}}
    \]
So, we have
    \[
        \tilde G=-\frac{\det \mathcal{M}}{2 \det \Omega }
    \]
Next we evaluate $\det \mathcal{M}$. By the usual matrix algebra (e.g., defining the determinant by a Gaussian integral and doing the ``$m-n$" integrals first),
\[
        \det \mathcal{M} = \det \pmatrix{
        \Omega & {\mathbf{v}} \cr
        {\mathbf{v}^{\mbox{T}}} & s \cr}
        =( {\det \Omega })\left[ {s-{\mathbf{v}}^{\mbox{T}}\Omega
        ^{-1}\mathbf{v}} \right]
\]
So we have the following expression for the Green function:
    \begin{equation} \label{eq8}
        \tilde{G}=-\frac{\det \mathcal{M}}{2 \det \Omega }
        =-\frac{1}{2} s + \frac{1}{2} {\mathbf{v}}^{\mbox{T}}\Omega ^{-1}{\mathbf{v}}
    \end{equation}

\section{V. Vacuum bubble}

To complete this general method of writing down the scattering
amplitude, we need to give the expression for the vacuum bubble
amplitude with a given topology $\mathscr{V}_{\mbox{M}}\left( T_a
\right)$ defined in equation (\ref{vacuum bubble}). This can always
be achieved by evaluating the bubble diagram by the second-quantized
method, but here we give a derivation by direct calculation in first-quantization. Note that the path integral (\ref{vacuum bubble}) is
the sum over all possible momentum configurations of the product of
the expectation values of the free evolution operator between two
states at the ends of each edge:
    \[
        \mathscr{V}_{\mbox{M}}\left( T_a \right)
        = \sum\limits_{\{p\}}\left\{\prod\limits_{a=1}^\mu { \left\langle p_a \left|
        e^{-T_a\left(p^2+m^2\right)/2} \right| p_a \right\rangle } \right\}
        = \sum\limits_{\{p\}}\left\{ \exp \left[ \sum\limits_{a=1}^\mu {-\frac{1}{2} T_a\left(p_a^2+m^2\right) }\right]
        \right\}
    \]
where $p_a$ is the momentum of the particle traveling on the $a$th
edge. The sum over all the configurations of $p_a$ can be written
as the the integration over all the possible values of $p_a$, but
they are not independent from each other. Each of the $\mu$ $p_a$'s
can be written as a linear combination of $L$ $k_i$'s where $L$ is
the number of loops of the graph and $k_i$ is the loop momentum on
the $i$th loop. So the amplitude $ \mathscr{V}_{\mbox{M}}\left( T_a
\right)$ can be written as
    \begin{equation}\label{eq11}
        \mathscr{V}_{\mbox{M}}\left( T_a \right)
        = \exp \left[ -\frac{1}{2} \left( \sum\limits_{a=1}^\mu { T_a } \right) m^2 \right]
        \int {\left( \prod\limits_{i}^{L}{dk_i} \right)\exp \left[ -\frac{1}{2} \underline{\sum\limits_{a=1}^\mu { T_a \left( \sum\limits_{i=1}^{L} {g_{ai}k_i} \right)^2 }}\right]}
    \end{equation}
where
    \[
        g_{ai} = \cases{
        1 & edge $a$ has the same direction as loop $i$ \cr
        -1 & edge $a$ has the opposite direction to loop $i$ \cr
        0 & edge $a$ is not on loop $i$ \cr}
    \]

It is easy to see the following points: (a) If edge $a$ is on loop
$i$, there is a $T_a k_i^2$ term in the underlined sum in equation
(\ref{eq11}), and vice versa. (b) If edge $a$ is on both loop $i$
and loop $j$, there is a term $2 T_a k_i k_j$ in the sum, and vice
versa. Further, if on edge $a$ both loops have the same (opposite)
direction, there is a positive (negative) sign before the term, and
vice versa. Thus if we use the factor $f(a,i,j)$ defined in equation
(\ref{eq9}), we can write the underlined part in equation
(\ref{eq11}) in a compact form, and further in terms of the period
matrix according to the definition (\ref{convenience}), or
definition (\ref{period matrix 2}) below:
    \[
        \sum\limits_{a=1}^\mu { T_a \left( \sum\limits_{i=1}^{L} {g_{ai}k_i} \right)^2 }
        =\sum\limits_{a=1}^\mu { \sum\limits_{i,j=1}^{L} { f\left(a,i,j\right) T_a k_i k_j } }
        =\sum\limits_{i,j=1}^{L} { \left[ \sum\limits_{a=1}^\mu{f\left(a,i,j\right) T_a } \right] k_i k_j }
        =\sum\limits_{i,j=1}^{L} { \Omega_{ij} k_i k_j }
    \]

The amplitude $\mathscr{V}_{\mbox{M}}\left( T_a \right)$ can then be
calculated easily:
    \begin{eqnarray}\label{determinant}
            \mathscr{V}_{\mbox{M}}\left( T_a \right) &=& \exp \left[ -\frac{1}{2} \left( \sum\limits_{a=1}^\mu { T_a } \right) m^2 \right]
            \int {\left( \prod\limits_{i}^{L}{dk_i} \right) \exp \left[ -\frac{1}{2} \sum\limits_{i,j=1}^{L} { \Omega_{ij} k_i k_j }
            \right]}\nonumber\\
            &=&  \exp \left[ -\frac{1}{2} \left( \sum\limits_{a=1}^\mu { T_a } \right) m^2
            \right] \left( \det \Omega \right) ^ {-D/2}
    \end{eqnarray}
where $D$ is the dimension of the spacetime.

\section{VI. Summary}

We now summarize our results for general amplitudes.  To find the Green function:

(1) Consider the ``circuit" (topology $\mbox{M}$ with two more
vertices at $\tau$ and $\tau'$ respectively) as a graph. Assign a
number to each edge. Define arbitrarily the direction of each edge.
Assign a number to each loop. Define arbitrarily the direction of
each loop.

(2) Find the period matrix $\Omega$ by the following definition:
    \begin{equation}\label{period matrix}
        \Omega_{ij}
        =\sum\limits_{k\,\in\,\rm{all\,edges}} f(k,i,j)r_k
    \end{equation}
where
\[
        f(n,i,j) = \cases{
        1 & $n$th edge is on both loops $i$ and $j$ with
        the same orientation \cr
        -1 & $n$th edge is on both loops $i$ and $j$ with
        opposite orientations \cr
         0 & $n$th edge is not on both loop $i$ and $j$ \cr}
\]

Each $r_k$ may be $\tau$, $\tau'$, a $T_a$ of the topological space $\mbox{M}$ without
external lines, $T_a-\tau$, or $T_a-\tau'$.
It is obvious that $\Omega$ does not depend on $\tau$ and $\tau'$, and only
depends on the  properties of $\mbox{M}$: To see this, note that the graph of the circuit has just two
more vertices (at positions $\tau$ and $\tau'$) than the graph of $\mbox{M}$ and each of them may seperate an edge in
the graph of $\mbox{M}$ into two ``sub-edges". But these changes on the
graph do not affect the period matrix: They neither give new loops nor
eliminate loops and the sub-edges will always be on or off a loop
simultaneously. Thus the period matrix of the graph of the circuit is
just the period matrix of the graph of $\mbox{M}$. And one only has to
calculate it once for one certain topology,

    \begin{equation}\label{period matrix 2}
        \Omega_{ij}
        =\sum\limits_{a=1}^\mu f(a,i,j)T_a
    \end{equation}

(3) Find a path (call it ``reference path") connecting $\tau$ and
$\tau'$. Choose its direction arbitrarily. Define the scalar $s$ as
the total resistance on the reference path.

(4) Find the vector $\mathbf{v}$ defined as follows:
    \[
        \mathbf{v}_i
        =\sum\limits_{k\,\in\,\rm{all\,edges}} f(k,i,0)r_k
    \]
where ``0" means the reference path.

(5) The Green function is given by
\[
        \tilde{G}=
        -\frac{1}{2} s + \frac{1}{2} {\mathbf{v}}^{\mbox{T}}\Omega ^{-1}{\mathbf{v}}
\]

The amplitude is then given by
\[
        \mathscr {A}\left( {\mbox{M},N} \right)=\prod\limits_{a=1}^\mu
        {\int_{\mbox{F}_a } {dT_a } } \, \mathscr{V}_{\mbox{M}}
        \prod\limits_{c\notin \mbox{C}}
        {\int_{\mbox{M}} {d\tau _c \exp \left[
        -\sum\limits_{i<j} {k_i \cdot k_j \tilde {G}\left( {\tau _i ,\tau _j }\right)} \right]} }
\]
with the Green function as above, and the vacuum bubble amplitude is
\[
            \mathscr{V}_{\mbox{M}}\left( T_a \right)
            =  \exp \left[ -\frac{1}{2} \left( \sum\limits_{a=1}^\mu { T_a } \right) m^2
            \right] \left( \det \Omega \right) ^ {-D/2}
\]

Although we have defined and derived all these quantities $s$,
$\mathbf{v}$ and $\Omega$ in terms of the concepts in the circuit
problem, it is easy to give the worldline geometric interpretation
by noting that the resistance is an analog of the proper time, i.e.,
the length of the worldline (Table \ref{tab1}). $s$ is just the
total proper time of the reference path, and the components of
$\mathbf{v}$ are the sums of the ``signed" proper time on the common
edges of the reference path and each loop. Entries of $\Omega$ are
sums of the ``signed" proper time on the common edges of each pair
of loops. (All the signs are given by $f$.) The components of
$\mathbf{v}$ and entries of $\Omega$ can also be expressed as
integrals of the ``Abelian differentials'' on the loops of the
worldline,
    \begin{eqnarray*}
        v_i &=& \int_{\tau'}^{\tau} { \omega_i }\\
        \Omega_{ij} &=& \ointctrclockwise_i \ { \omega_j }
    \end{eqnarray*}
where $\omega_i$ is the line element on loop $i$ and the second
integral is around loop $i$ along its direction.
Then our expression for the particle amplitude has a similar structure to the bosonic string amplitude, with the Green function on
the 2D worldsheet \cite{D'Hoker:1988ta}:
    \[
        \tilde {G} \left( w , z \right) = - 2 \ln \left| E\left( w,z \right) \right| + 2 \pi\ \text{Im} \int_w^z \mathbf{\omega} \left( \text{Im}\ \Omega \right)^{-1} \text{Im} \int_w^z
        \mathbf{\omega}
    \]
where $E$ is the prime form, the vector $\mathbf{\omega}$ contains
the basis of the Abelian differentials and the matrix $\Omega$ is
the period matrix.

\section{VII. Examples}

\begin{figure}
        \includegraphics{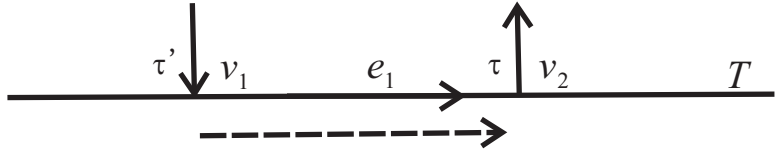}
        \caption{\label{fig4}
        The topology of a line with length $T$. There is no loop and hence no
        period matrix nor vector $\mathbf{v}$. The only path between $\tau$
        and $\tau'$ is the edge connecting the two vertices $e_1$, so we
        choose it as the reference path.}
    \end{figure}

Here we give some examples of obtaining the Green functions on
different topologies. For the finite line of length $T$ in Fig.\
\ref{fig4}, there is no loop, so there is no period matrix $\Omega$
nor vector $\mathbf{v}$. $s$ is just the total resistance between
$\tau$ and $\tau'$. So the Green function is
    \[
        \tilde{G}_{-} \left( \tau , \tau' \right)=-\frac{1}{2}s=-\frac{1}{2}\left|{\tau-\tau'}\right|
    \]
and
    \[
        \mathscr{V}_{-} \left( T \right)=  \exp \left( -\frac{1}{2} T m^2 \right)
    \]
The amplitude is then given by equation (\ref{eq10}). One just has
to note that there is one modulus $T$ for this case and to fix the
residual symmetry:  Two of the external lines should be fixed at one
end of the line and another two should be fixed at the other end.

    \begin{figure}
        \includegraphics{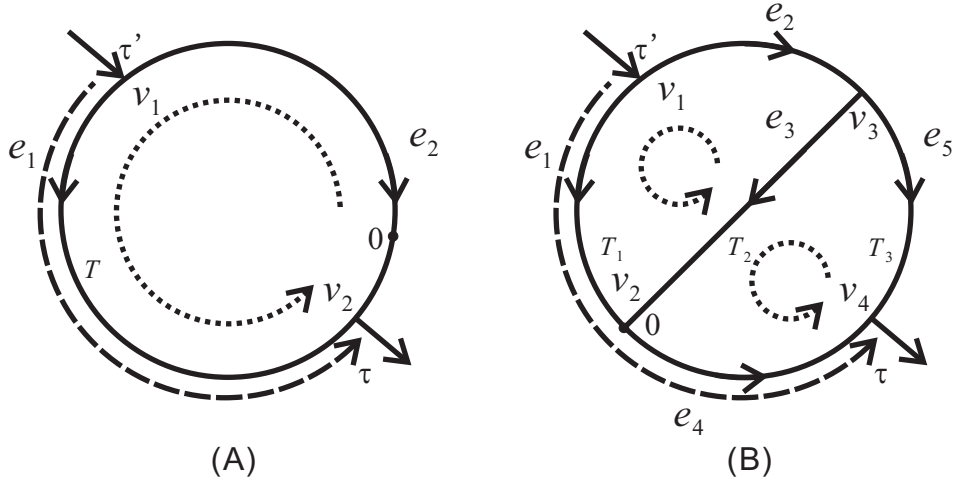}
        \caption{\label{fig5}
        (A) The one-loop topology.
        $T$ is the circumference of the circle.
        There are two edges ($e_1$,$e_2$) and two vertices ($v_1$,$v_2$).
        $\tau$ and $\tau'$ are the lengths from $v_1$ and $v_2$ to the origin though a counterclockwise path.
        The directions of the edges are chosen arbitrarily and marked in the figure.
        There is only one loop and it is marked by dotted lines.
        The reference path is marked by a dashed line.
        (B) A two-loop topology.
        $T_1$, $T_2$ and $T_3$ are the lengths of the three arcs.
        $\tau$ and $\tau'$ are respectively the length on the 3rd and 1st arc from the origin.
        (If the Green function with $\tau$ and $\tau'$ on different arcs is needed, simply repeat the steps for this special case.)
        There are 5 edges ($e_1$ to $e_5$) and 4 vertices ($v_1$ to $v_4$).
        The directions of the edges are chosen arbitrarily and marked in the figure.
        There are two independent loops and they are marked by dotted lines.
        The reference path is marked by a dashed line.}
    \end{figure}
For the circle, there is one loop as shown in Fig.\ \ref{fig5}(A). So
the period matrix is $1 \times 1$. The only entry of $\Omega$ is
then
    \[
        \Omega = T
    \]
And according to the definition, $s$ and $\mathbf{v}$ are
    \[
        s = \left|{\tau-\tau'}\right|
    \]
    \[
        \mathbf{v} = \left|{\tau-\tau'}\right|
    \]
Thus the Green function on the circle is
    \[
        \tilde{G}_{\bigcirc} \left( \tau , \tau' \right) =-\frac{1}{2} s + \frac{1}{2} {\mathbf{v}}^{\mbox{T}}\Omega ^{-1}{\mathbf{v}}
        =-\frac{1}{2}\left|{\tau-\tau'}\right| + \frac{(\tau-\tau')^2}{2T}
    \]
and the vacuum bubble amplitude is
    \[
        \mathscr{V}_{\bigcirc}\left( T \right)
        =  \exp \left( -\frac{1}{2} T m^2 \right) T ^ {-D/2}
    \]

For the Green function on the 2-loop graph shown in Fig.\
\ref{fig5}(B), the period matrix is $2\times2$ and $\mathbf{v}$ is a
two-component vector. $\Omega$, $s$ and $\mathbf{v}$ are
\begin{eqnarray*}
        \Omega &=&\pmatrix{
        T_1 +T_2  & -T_2 \cr
        -T_2  & T_2 +T_3 \cr}\\
        s&=&\tau'+\tau\\
        \mathbf{v}&=&( \tau',\tau )^{\mbox{T}}
\end{eqnarray*}
So the Green function is, by plugging all the above into equation
(\ref{eq8}),
    \[
        \tilde {G}_{\ominus} \left( \tau , \tau' \right) =-\frac{T_1 \left( {T_2 +T_3 -\tau } \right)\tau +T_1
        \left( {T_2 +T_3 } \right){\tau }'-T_3 {\tau }'^2+T_2 \left( {T_3
        -\tau -{\tau }'} \right)\left( {\tau +{\tau }'} \right)}{2\left(
        {T_1 T_2 +T_2 T_3 +T_3 T_1 } \right)}
    \]
and the vacuum bubble amplitude is
    \[
        \mathscr{V}_{\ominus}\left( T_1,T_2,T_3 \right)
        =  \exp \left[ -\frac{1}{2} \left( T_1 + T_2 + T_3 \right) m^2 \right] \left( T_1 T_2 + T_2 T_3 + T_3 T_1 \right) ^ {-D/2}
    \]

\section{VIII. Conclusions}

Based on the first-quantized formalism, the discussion in this
paper gives a new method to evaluate the scattering amplitude of
scalar field theory at arbitrary loop level. (The procedure was given in section VI.) The form of the
Green function is shown to be similar to that on the worldsheet in
bosonic string theory.
The method applies not only to 1PI graphs, but arbitrary graphs that might appear in S-matrices.

The amplitude obtained by the first-quantized method in this paper
easily can be shown to be equivalent to the amplitude from second-quantization. Further, the singularities of general diagrams can be
discussed and the Landau conditions can be obtained. The occurrence
of a singularity on the physical boundary was shown to be related to
a sequence of interactions connected by real particle paths by
Coleman and Norton \cite{Coleman:1965xm}. Here this same picture
emerges more naturally: The Feynman diagram is interpreted as
particle path (through background fields) and the integral over the
proper time is introduced from the beginning.

The extension to diagrams with spinning particles will be the next
thing to understand. It is expected to be parallel to Strassler's
discussion on various theories \cite{Strassler:1992zr}, with the
Green function on a circle replaced by those on other topologies.
However, although the 2-loop Green function has been known for a
long time, no full-fledged extension to even 2-loop Yang-Mills
theory has yet been found, because of the complexity and difficulty
entering at multi-loop level. Previous attempts include
generalization from Bern and Kosower's string approach
\cite{DiVecchia:1995in}\cite{DiVecchia:1996uq} and Strassler's
particle approach \cite{Sato:1998sf}\cite{Sato:2000cr}.

It is now clear that scalar field theory can be seen as the limit of
the bosonic string theory, by suppressing the length of the string.
Can the bosonic string worldsheet be seen as the sum of all the
Feynman diagrams of the scalar field theory that are random lattices
of the worldsheet? This is another interesting question that may be
related to the result in this paper.

\section*{Acknowledgment}

This work was supported in part by National Science Foundation
Grant No.\ PHY-0354776.

\end{document}